\documentclass[conference]{IEEEtran}

\usepackage[pdftex]{graphicx}

\usepackage[T1]{fontenc}
\usepackage{babel,blindtext}
\usepackage{subcaption}
\usepackage{color}
\usepackage{varioref}
\usepackage{textcomp}
\usepackage{amsthm}
\usepackage{amsmath}
\usepackage{amssymb}
\usepackage{graphicx}
\usepackage{setspace}
\usepackage{esint}
\usepackage{algorithm}
\usepackage{algpseudocode}
\usepackage{pifont}
\usepackage{wrapfig}
\usepackage{multirow}
\usepackage{tabu}
\usepackage{amsmath}

\usepackage{enumitem}
\usepackage{cite}
\usepackage{cleveref}
\usepackage{dsfont}

\makeatletter

\newcommand{\Rmnum}[1]{\expandafter\@slowromancap\romannumeral #1@}
\makeatother

\newtheorem{remark}{Remark}

\theoremstyle{definition}

\providecommand{\propositionname}{Proposition}

\ifCLASSINFOpdf

\else

\fi

\usepackage[T1]{fontenc}
\usepackage{etoolbox}

\makeatletter
\patchcmd{\maketitle}{\@fnsymbol}{\@alph}{}{}  
\makeatother

\title{Federated Learning with Downlink Device Selection}
\author{\IEEEauthorblockN{Mohammad Mohammadi Amiri, Sanjeev R. Kulkarni, H. Vincent Poor}
\IEEEauthorblockA{Department of Electrical and Computer Engineering, Princeton University, Princeton, NJ 08544, USA}
}

\begin{document}

\maketitle

\begin{abstract}
We study federated edge learning, where a global model is trained collaboratively using privacy-sensitive data at the edge of a wireless network.
A parameter server (PS) keeps track of the global model and shares it with the wireless edge devices for training using their private local data. 
The devices then transmit their local model updates, which are used to update the global model, to the PS.  
The algorithm, which involves transmission over PS-to-device and device-to-PS links, continues until the convergence of the global model or lack of any participating devices.
In this study, we consider device selection based on downlink channels over which the PS shares the global model with the devices.
Performing digital downlink transmission, we design a partial device participation framework where a subset of the devices is selected for training at each iteration. 
Therefore, the participating devices can have a better estimate of the global model compared to the full device participation case which is due to the shared nature of the broadcast channel with the price of updating the global model with respect to a smaller set of data.
At each iteration, the PS broadcasts different quantized global model updates to different participating devices based on the last global model estimates available at the devices.
We investigate the best number of participating devices through experimental results for image classification using the MNIST dataset with biased distribution. \makeatletter{\renewcommand*{\@makefnmark}{}\footnotetext{This work was supported by the U.S. National Science Foundation under Grants CCF-0939370 and CCF-1908308.}\makeatother}
\end{abstract}


\section{Introduction}\label{SecIntro}



With the growing number of Internet of Things (IoT) devices and services, huge amounts of data are being generated at the wireless edge.
We also observe a rapid increase in the Artificial Intelligence (AI)-driven applications deploying increasingly complex machine learning (ML) models to gather intelligence from data.
While this is traditionally carried out by a cloud server, offloading such massive amounts of data at the wireless network edge to a cloud server is prohibitive in terms of communications cost and may violate privacy concerns.  
Instead, a new paradigm, referred to as \textit{federated learning} (FL), offers pushing the network intelligence to the edge by elevating edge computing capabilities in distributed networks.
FL allows distributed devices to train a global ML model collaboratively with the help of a parameter server (PS) without sharing the raw data \cite{DCKonecnyFederated}.

With FL, the PS keeps track of the global model and shares it with the devices for training using their local data. 
After training the global model, the devices transmit their updates to the PS, which uses them to update the global model.
FL comes with its own challenges in practical settings including communications over unreliable links in both PS-to-device and device-to-PS directions.
Several studies have addressed this challenge through limiting the communication requirements, particularly in the device-to-PS direction \cite{DCKonecnyFederated,McMahan2017CommunicationEfficientLO,KonecnyFLBeyondData}.
However, these studies ignore the underlying communication channels and consider rate-limited bit-pipes as communication links.

Recently a new line of research, namely \textit{federated edge learning} (FEEL), has been created considering the physical layer characteristics of the underlying wireless network in the FL framework \cite{MohammadDenizDSGDCS,FLTWCMohammadDenizFading,KaibinParallelWork,YangFedLearOverAirComp}.  
Several studies have employed over-the-air computation by using the superposition property of the multiple access channel (MAC) in the device-to-PS direction and have shown its advantages in guaranteeing reliable communications of the updates from the devices to the PS \cite{MohammadDenizDSGDCS,KaibinParallelWork,YangFedLearOverAirComp,FLTWCMohammadDenizFading,MohTolDenSanjVinceBlindFL,CohenAnalogGDDL,RaviCommEffFLGaussian}. 
FEEL should be tolerant to the drop of any devices during the training.  
This arises from the fact that the devices share the channel resources, and typically a huge number of devices can be potentially used for training, in which case each device may have access to only a limited amount of channel resources for transmission.
Also, wireless devices may stop participating in the training at any time for various reasons including lack of power, lack of network coverage, and privacy concerns.
Therefore, several studies consider FEEL with partial device participation sharing the limited resources for various selection techniques \cite{YangArafaVinceAgeBasedFL,YuxuanFLRedundantData,HowardVinceSchedulingsFL,FLTWCConvergenceMohammadDenizSanjVince}.

Recent studies have focused on noisy PS-to-device channel, as well as transmission of a compressed global model to the devices to reduce the communication footprint \cite{MohammadFLConvNoisyDownlink,JinHyunAhnFLNoisyDownlink,ExpandingRedClientResFL,DoubleSqueezeTangConf,mohammadDenizSanjVinceQuanUpdates}. 
Papers \cite{JinHyunAhnFLNoisyDownlink,MohammadFLConvNoisyDownlink} investigated analog transmission of the global model to the devices with \cite{MohammadFLConvNoisyDownlink} providing convergence guarantees.  
Compressing the global model at the PS may lead to a coarse estimate, which is due to the high average-to-peak ratio of the entries of global model vector and its high empirical variance \cite{mohammadDenizSanjVinceQuanUpdates}.
To overcome this challenge, \cite{ExpandingRedClientResFL} employs a random linear projection of the global model at the PS before the compression, \cite{DoubleSqueezeTangConf} performs error compensation before compressing the global model, and \cite{mohammadDenizSanjVinceQuanUpdates} introduces compression of the global model updates with respect to the last global model estimate available at the devices, which is shown to outperform the other two approaches. 

In this paper, we build upon our approach in \cite{mohammadDenizSanjVinceQuanUpdates} and focus on digital transmission of the global model updates at the PS.
We extend the scheme in \cite{mohammadDenizSanjVinceQuanUpdates} to the partial device participation scenario by considering transmission of different messages to different devices in the PS-to-device direction, where the devices are selected for participation in each training iteration based on their downlink channels.
Specifically, the PS transmits compressed global model updates to the participating devices based on the last global model estimates available at the devices where the rates are set to be within the capacity region of the underlying wireless fading broadcast channel from the PS to the devices.
The number of participating devices directly impacts the performance of FEEL, since the devices share the limited channel resources, and fewer participating devices leads to better estimates of the information exchanged in the network between different nodes. 
On the other hand, fewer participating devices reduces the data usage in each training iteration. 
Numerical results show that partial device participation based on the downlink channels can significantly improve the performance of FEEL.

\textit{Notation}: We denote the set of real and complex numbers by $\mathbb{R}$ and $\mathbb{C}$, respectively. 
We let $[i] \triangleq \{ 1, \dots, i \}$. 
Vectors of length $i$ with all zero and one entries are denoted by $\boldsymbol{0}_i$ and $\boldsymbol{1}_i$, respectively. 
We denote a circularly symmetric complex Gaussian distribution with real or imaginary component with variance $\sigma^2/2$ by $\mathcal{C N} \left( 0,\sigma^2 \right)$.
We represent the cardinality of a set or the magnitude of a complex value by $\left| \cdot \right|$, and the $l_2$ norm of a vector $\boldsymbol{x}$ by $\left\| \boldsymbol{x} \right\|_2$. 
For a set $\cal S$ of real values, ${\rm{indmax}}_{K} \mathcal{S}$ returns a $K$-element set of indices of $\mathcal{S}$ with the largest values. 
We also let
$(a)^+ \triangleq \max \{a, 0\}$, and for any event $A$, $\mathds{1}(A)=1$, if event $A$ is true, and $\mathds{1}(A)=0$, otherwise.



\section{System Model}\label{SecProbFormul}

We consider FEEL across $M$ devices, where device $m$ has access to dataset $\mathcal{B}_m$ of size $B_m = \left| \mathcal{B}_m \right|$.
The devices use their local datasets to train a global model $\boldsymbol{\theta} \in \mathbb{R}^d$ with the help of a parameter server (PS) in order to minimize loss function $F \left( \boldsymbol{\theta} \right) = \sum\nolimits_{m=1}^{M} \frac{B_m}{B} F_m(\theta)$,
where $B \triangleq \sum\nolimits_{m=1}^{M} B_m$, and $F_m \left( \boldsymbol{\theta} \right) = \frac{1}{B_m} \sum\nolimits_{\boldsymbol{u} \in \mathcal{B}_m} f \left(\boldsymbol{\theta}, \boldsymbol{u} \right)$ is the loss function at device $m$ with $f$ denoting an empirical loss function. 

\subsection{FEEL System}
Minimization of loss function $F$ is carried out iteratively using the local datasets at the devices with the PS having access to the global model. 
At each iteration, the global model is shared with the devices and updated locally using the data at the devices.
The devices then transmit their local model updates to the PS responsible for updating the global model.  
The algorithm continues until the convergence of the global model, or lack of participating devices for training.

The downlink (PS-to-device) and uplink (device-to-PS) connections are wireless, and a noisy transmission over wireless medium takes place during each downlink and uplink communication round. 
During iteration $t$, the PS shares the global model $\boldsymbol{\theta} (t)$ with the devices, and they perform a $\tau$-step stochastic gradient descent (SGD) to update the global model using their local data.
Let $\widehat{\boldsymbol{\theta}}_m (t)$ denote the noisy estimate of the global model $\boldsymbol{\theta} (t)$ at device $m$, $m \in [M]$.
For the $i$-th step SGD algorithm, device $m$ selects a random subset of its local data, denoted by $\boldsymbol{\xi}_m^i (t)$, and updates the model as follows:
\begin{align}\label{EQTauStepSGD}
\boldsymbol{\theta}_m^{i+1} (t) = \boldsymbol{\theta}_m^i (t) - \eta^i_m (t) \nabla F_m \left( \boldsymbol{\theta}_m^i (t), \boldsymbol{\xi}_m^i (t)  \right),\, \mbox{$i \in [\tau]$},   
\end{align}
where $\boldsymbol{\theta}_m^1 (t) = \widehat{\boldsymbol{\theta}}_m (t)$, $\eta^i_m (t)$ denotes the learning rate, and $\nabla F_m \left( \boldsymbol{\theta}_m^i (t), \boldsymbol{\xi}_m^i (t)  \right)$ is the stochastic gradient which provides an unbiased estimate of the true gradient $\nabla F_m \left( \boldsymbol{\theta}_m^i (t)  \right)$.
Device $m$ sends its model update $\Delta \boldsymbol{\theta}_m (t) = \boldsymbol{\theta}_m^{\tau+1} (t) - \boldsymbol{\theta}_m^{1} (t)$ to the PS, computed after performing the $\tau$-step SGD, where we denote the estimate of $\Delta \boldsymbol{\theta}_m (t)$ at the PS by $\Delta \widehat{\boldsymbol{\theta}}_m (t)$.
The PS then updates the global model as follows:
\begin{align}
\boldsymbol{\theta} (t+1) =\boldsymbol{\theta} (t) + \frac{1}{M} \sum\nolimits_{m=1}^{M} \Delta \widehat{\boldsymbol{\theta}}_m \left( t \right).   
\end{align}

FEEL needs to address the partial device participation case, when only a subset of the devices participate in the training at each iteration.
First, we need to consider the fact that some of the wireless devices might stop participating in the training for various reasons.
Furthermore, due to the limited wireless resources shared among the participating devices, the information that can be conveyed using each link may be limited after allocating the channel resources. 
In this paper, we adopt partial device participation, where at iteration $t$ only a subset of devices $\mathcal{K} (t) \subset [M]$ with size $K = \left| \mathcal{K} (t) \right|$ is chosen by the PS to participate in the training. 
Accordingly, the PS updates the global model after receiving the local model updates from the participating devices as
\begin{align}
\boldsymbol{\theta} (t+1) =\boldsymbol{\theta} (t) + \frac{1}{K} \sum\nolimits_{k \in \mathcal{K} (t)} \Delta \widehat{\boldsymbol{\theta}}_k \left( t \right).   
\end{align}



\subsection{Wireless Medium}\label{SubSecWirelessMed}
Each iteration includes two communication phases over downlink, where the PS shares the global model with the devices, and uplink, where the participating devices transmit their local updates to the PS.
Here we model the downlink and uplink communication channels.

\noindent \textbf{Downlink channel.}
We assume that the downlink is a parallel fading broadcast channel with $s^{\rm{dl}}$ sub-channels.
We denote the channel gains vector from the PS to device $m$ at iteration $t$ by $\boldsymbol{h}^{\rm{dl}}_m (t) = [{h}^{\rm{dl}}_{m, 1} (t), ..., {h}^{\rm{dl}}_{m, {s^{\rm{dl}}}} (t)] \in \mathbb{C}^{s^{\rm{dl}}}$ with each entry independent and identically distributed (iid) according to $\mathcal{CN} (0, (\sigma^{\rm{dl}})^2)$, and the additive noise vector at device $m$ by $\boldsymbol{z}^{\rm{dl}}_m (t) \in \mathbb{C}^{s^{\rm{dl}}}$ with each entry iid according to $\mathcal{CN} (0, 1)$.
We assume that the PS has channel state information (CSI) about the downlink, and each device has the CSI of its link, and we impose an average power constraint $P^{\rm{dl}}$ at the PS.

\noindent \textbf{Uplink channel.}
We model the uplink channel with a parallel fading multiple access channel (MAC) with $s^{\rm{ul}}$ sub-channels.
The channel gains vector from device $m$ to the PS at iteration $t$ is denoted by $\boldsymbol{h}^{\rm{ul}}_m (t) = [{h}^{\rm{ul}}_{m, 1} (t), ..., {h}^{\rm{ul}}_{m, {s^{\rm{ul}}}} (t)] \in \mathbb{C}^{s^{\rm{ul}}}$ with each entry iid according to $\mathcal{CN} (0, (\sigma^{\rm{ul}})^2)$. 
We also denote the additive noise vector at the PS by $\boldsymbol{z}^{\rm{ul}}_m (t) \in \mathbb{C}^{s^{\rm{ul}}}$ with each entry iid according to $\mathcal{CN} (0, 1)$.
The channel input at each device is limited to an average transmit power constraint $P^{\rm{ul}}$, and the PS has full CSI, while each device knows its CSI.

\section{Communications Protocol}
We consider digital communications of the information from the PS and the devices, where each node compresses its message and transmits it at a rate below its channel capacity. 
We first present the channel capacities of the downlink and uplink, and then introduce the quantization technique used for compression. 
For the ease of presentation, we drop the dependency of the channels on the iteration count $t$.

\subsection{Downlink Channel Capacity}
Recall that the downlink is a parallel fading broadcast channel with $s^{\rm{dl}}$ sub-channels.
Let us consider serving an arbitrary subset of the devices $\mathcal{K} \subset [M]$ of size $K$.
First, we focus on a fading broadcast channel with a single sub-channel, i.e., $s^{\rm{dl}}=1$, where the PS serves the $K$ devices.
We denote the channel gain from the server to device $k$ by ${h}^{\rm{dl}}_k$ and the channel gains vector by $\boldsymbol{h}^{\rm{dl}}$. 
This yields a degraded broadcast channel with a well known capacity \cite{BergmansDegradedBCCapacity}.
Without loss of generality, let $\mathcal{K} = [K]$, and $| {h}^{\rm{dl}}_1 | \le | {h}^{\rm{dl}}_2 | \le \cdots \le | {h}^{\rm{dl}}_K |$. 
For some non-negative values $\alpha_k$, $\forall k \in \mathcal{K}$, such that $\sum\nolimits_{k=1}^{K} \alpha_k = 1$, the boundary of the capacity region $\mathcal{C}^{\rm{dl}}_{\mathcal{K}, b} (\boldsymbol{h}^{\rm{dl}}, P^{\rm{dl}})$ is given by
\begin{align}
& \mathcal{C}^{\rm{dl}}_{\mathcal{K}, b} (\boldsymbol{h}^{\rm{dl}}, P^{\rm{dl}}) = \Big\{ [C_1, ..., C_K]^T: \nonumber\\
& \; C_k = \log_2 \Big( 1+\frac{\alpha_k \left| {h}^{\rm{dl}}_k \right|^2 P^{\rm{dl}}}{1+ \sum\nolimits_{j=k+1}^{K} \alpha_j \left| {h}^{\rm{dl}}_j \right|^2 P^{\rm{dl}}} \Big), \; k \in \mathcal{K} \Big\}.    
\end{align}
Accordingly, the capacity region of the parallel fading broadcast channel is given by \cite{ParallelGaussianBCTse} 
\begin{align}
\mathcal{C}^{\rm{dl}}_{\mathcal{K}} (P^{\rm{dl}}) = \bigcup\limits_{\{P^{\rm{dl}}_1, ..., P^{\rm{dl}}_{s^{\rm{dl}}}\}:\sum\limits_{i=1}^{s^{\rm{dl}}} P^{\rm{dl}}_i = P^{\rm{dl}}} \sum\limits_{i=1}^{s^{\rm{dl}}} \mathcal{C}^{\rm{dl}}_{\mathcal{K}, b} (\boldsymbol{h}^{\rm{dl}}_i, P^{\rm{dl}}_i),   
\end{align}
where $\boldsymbol{h}^{\rm{dl}}_i \triangleq [{h}^{\rm{dl}}_{1,i}, ..., {h}^{\rm{dl}}_{K,i}]$, for $i \in [s^{\rm{dl}}]$.
The optimal rate and power allocation across sub-channels to maximize the sum-capacity is given by \cite{ParallelGaussianBCTse}
\begin{subequations}
\begin{align}
&C^{\rm{dl}}_{k, i} =  \begin{cases}
\log_2 \Big( 1 + \Big( \frac{| h_{k, i}^{\rm{dl}}|^2}{\lambda} -1\Big)^+ \Big), & \mbox{if $| h_{k, i}^{\rm{dl}}| = \max\limits_{k' \in \mathcal{K}} | h_{k', i}^{\rm{dl}}|$},\\
0, & \mbox{otherwise},
\end{cases} \\
& P^{\rm{dl}}_{i} = \Big( \frac{1}{\lambda} - \frac{1}{\max\limits_{k \in \mathcal{K}} | h_{k, i}^{\rm{dl}}|^2} \Big)^+,
\end{align}
\end{subequations}
where $\lambda$ is chosen to satisfy the power constraint $\sum\nolimits_{i=1}^{s^{\rm{dl}}} {P}_{i}^{\rm{dl}} = {P}^{\rm{dl}}$,
and the capacity of the channel to device $k$ is given by $C_{\mathcal{K}, k}^{\rm{dl}} = \sum\nolimits_{i=1}^{s^{\rm{dl}}} C^{\rm{dl}}_{k, i}$, $k \in \mathcal{K}$.
The optimal solution provided above dictates that in each sub-channel the information for no more than a single user, the one with the best channel quality in that sub-channel, is broadcast.
Also, the power allocation across different sub-channels follows a water-filling solution.

\subsection{Uplink Channel Capacity}
The uplink is modelled by a parallel fading MAC with $s^{\rm{ul}}$ sub-channels. 
We first consider $s^{\rm{ul}}=1$, where $K$ devices in $\mathcal{K}$ share the wireless medium to communicate with the PS.
We denote the channel gains vector by $\boldsymbol{h}^{\rm{ul}} = [{h}^{\rm{ul}}_1, ..., {h}^{\rm{ul}}_K]$, where ${h}^{\rm{ul}}_k$ is the channel gain from device $k$ to the PS. 
For an average power $P^{\rm{ul}}_k$ at device $k$ and $\boldsymbol{P}^{\rm{ul}} \triangleq [P_1^{\rm{ul}}, ..., P_K^{\rm{ul}}]$, the capacity region of the above fading MAC is given by
\begin{align}
& \mathcal{C}^{\rm{ul}}_{\mathcal{K}, b} (\boldsymbol{h}^{\rm{ul}}, \boldsymbol{P}^{\rm{ul}}) =\Big\{ [C_1, ..., C_K]^T:\nonumber\\
& \sum\nolimits_{k \in \mathcal{S}} C_k \le \log_2 \Big( 1 + \sum\nolimits_{k \in \mathcal{S}} \left| {h}^{\rm{ul}}_k \right|^2 P^{\rm{ul}}_k \Big),\; \forall \mathcal{S} \subset \mathcal{K} \Big\}.  
\end{align}
The capacity region of the parallel fading MAC is then given by \cite{ParallelGausianMACCapTse}
\begin{align}
\mathcal{C}^{\rm{ul}}_{\mathcal{K}} (P^{\rm{ul}}) = \bigcup\limits_{\quad \scriptstyle\{ P_{k,1}^{\rm{ul}},...,P_{k,{s^{\rm{ul}}}}^{\rm{ul}}\} :\hfill\atop
\scriptstyle\sum\nolimits_{i = 1}^{{s^{\rm{ul}}}} {P_{k,i}^{\rm{ul}}}  = {P^{\rm{ul}}},\forall k \in \mathcal{K}\hfill} {\sum\limits_{i = 1}^{{s^{\rm{ul}}}} } \mathcal{C}^{\rm{ul}}_{\mathcal{K}, b} (\boldsymbol{h}^{\rm{ul}}_i, \boldsymbol{P}_i^{\rm{ul}}),    
\end{align}
where $\boldsymbol{h}^{\rm{ul}}_i \triangleq [{h}^{\rm{ul}}_{1,i}, ..., {h}^{\rm{ul}}_{K,i}]$ and $\boldsymbol{P}^{\rm{ul}}_i \triangleq [{P}^{\rm{ul}}_{1,i}, ..., {P}^{\rm{ul}}_{K,i}]$, $i \in [s^{\rm{ul}}]$.
Assuming a total power constraint $\sum\nolimits_{k=1}^{K} \sum\nolimits_{i=1}^{s^{\rm{ul}}} P_{k,i}^{\rm{ul}} = K P^{\rm{ul}}$, the optimal power allocation across sub-channels to maximize the sum-capacity is given by \cite{ParallelGausianMACCapTse}
\begin{align}
P_{k,i}^{\rm{ul}} = 
\begin{cases}
\Big( \frac{1}{\lambda} - \frac{1}{\left| h_{k, i}^{\rm{ul}}\right|^2} \Big)^+, & \mbox{if $| h_{k, i}^{\rm{ul}}| = \max\limits_{k' \in \mathcal{K}} | h_{k', i}^{\rm{ul}}|$},\\
0, & \mbox{otherwise},
\end{cases}
\end{align}
where $\lambda$ is chosen to satisfy $\sum\nolimits_{k=1}^{K} \sum\nolimits_{i=1}^{s^{\rm{ul}}} P_{k,i}^{\rm{ul}} = K P^{\rm{ul}}$. 
Accordingly, the sum capacity is given by
\begin{align}
C^{\rm{ul}}_{\mathcal{K}, \rm{sum}} = \sum\nolimits_{i=1}^{s^{\rm{ul}}} \log_2 \Big( 1 + \Big( \frac{\max\nolimits_{k \in \mathcal{K}} |  h_{k, i}^{\rm{ul}}|^2}{\lambda} -1\Big)^+ \Big),   
\end{align}
and for a symmetric rate allocation, the capacity of each device is given by $C_{\mathcal{K}, k}^{\rm{ul}} = C^{\rm{ul}}_{\mathcal{K}, \rm{sum}}/K$, $\forall k \in \mathcal{K}$.
Similar to the downlink transmission, with the uplink transmission, at each sub-channel the power is allocated to only a single device with the best channel condition in that sub-channel.
We note that, besides its CSI, each device needs to know the indices of the sub-channels that it has the best channel conditions among all the devices (the sub-channels that it uses for transmission), and we assume that this information is provided by the PS.

\begin{remark}
For the above capacity analysis, we have assumed slow fading and coding across several blocks with invariant channel gains. 
This provides an upper bound on the capacity of the downlink and uplink channels presented in Section \ref{SubSecWirelessMed}. 
\end{remark}

\subsection{Quantization Technique}
We follow the stochastic quantization technique presented in \cite{mohammadDenizSanjVinceQuanUpdates}.
Let $\boldsymbol{x} \in \mathbb{R}^d$ be a vector with the $i$-th component denoted as $x_i$. 
We define $x_{\rm{max}} \triangleq \max \{ |\boldsymbol{x}| \}$ and $x_{\rm{min}} \triangleq \min \{ |\boldsymbol{x}| \}$.
For a quantization level $q \ge 1$, we define, for $i \in [d]$,
\begin{align}
& {Q} \left({x}_i, q\right) \triangleq {\rm{sign}} \left( {x}_i \right) \cdot \Big( x_{\rm{min}} \nonumber\\
& \qquad \qquad \quad + \left(x_{\rm{max}} - x_{\rm{min}} \right) \cdot \varphi \Big( \frac{\left| {x}_i \right| - x_{\rm{min}}}{x_{\rm{max}} -x_{\rm{min}}}, q \Big) \Big),    
\end{align}
where $\varphi$ is a stochastic quantization function defined next. 
Let $x \in \mathbb{R}$ such that $0 \le x \le 1$ and $q \ge 1$, and $l \in \{ 0, 1, ..., q-1\}$ be an integer such that $x \in [l/q, (l+1)/q)$. 
We define
\begin{align}
\varphi \left( x, q \right) \triangleq \begin{cases} 
l / q, & \mbox{with probability $1 - x q + l$},\\
(l+1) / q, & \mbox{with probability $x q - l$}.
\end{cases}
\end{align}
We further define the vector quantization as $\boldsymbol{Q}(\boldsymbol{x}, q) \triangleq [Q({x}_1, q), \cdots, Q({x}_d, q) ]^T$.
With this quantization approach, each vector $\boldsymbol{x}$ is represented by a total number of
\begin{align}
R_q = 64 + d \left( 1 + \log_2(q+1) \right) \mbox{ bits}.    
\end{align}

\section{Downlink Device Selection Technique}
Here we present our FEEL approach with device selection in the downlink. 
We extend our work in \cite{mohammadDenizSanjVinceQuanUpdates} by allowing the PS to broadcast different messages to different devices that enables partial device participation at each training iteration.   
At each iteration, the PS selects a subset of the devices and broadcasts different versions of the global model to the selected devices with a rate vector within the capacity region $\mathcal{C}^{\rm{dl}}(P^{\rm{dl}})$.
The global model estimate at device $m$ at iteration $t$ is denoted by $\widehat{\boldsymbol{\theta}}_m (t)$, and if $m \in \mathcal{K} (t)$, i.e., device $m$ is selected to participate in the training at iteration $t$, it performs the $\tau$-step SGD algorithm.
It then quantizes the local model update $\Delta {\boldsymbol{\theta}}_m \left( t \right)$, along with error compensation, and transmits it to the PS with a rate no larger than $C^{\rm{ul}}_m$, $m \in [M]$. 
In the following, we elaborate the downlink and uplink transmission.

\subsection{Downlink Transmission}
At iteration $t$, the PS first selects $K$ devices with the largest $\sum\nolimits_{i=1}^{s^{\rm{dl}}} \big\| \boldsymbol{h}^{\rm{dl}}_m (t) \big\|^2_2$, $\forall m \in [M]$; that is,
\begin{align}
\mathcal{K} (t) = {\rm{indmax}}_{K} \Big\{ \sum\nolimits_{i=1}^{s^{\rm{dl}}} \big\| \boldsymbol{h}^{\rm{dl}}_1 (t) \big\|^2_2, ..., \sum\nolimits_{i=1}^{s^{\rm{dl}}} \big\| \boldsymbol{h}^{\rm{dl}}_M (t) \big\|^2_2 \Big\}.    
\end{align}
The PS aims to transmit $\boldsymbol{\theta} (t) - \widehat{\boldsymbol{\theta}} (t_k)$ to device $k \in \mathcal{K}(t)$ over the downlink, where $t_k$ denotes the iteration index during which device $k$ was selected last time, and it is initialized with $t_k = 0$. 
We will show that, with the proposed approach, the PS knows about $\widehat{\boldsymbol{\theta}} (t_k)$ as side information and can remove this information from the current global model $\boldsymbol{\theta} (t)$, $\forall k \in \mathcal{K} (t)$.

For a set of selected devices $\mathcal{K}(t)$ with the downlink capacity region $\mathcal{C}^{\rm{dl}}_{\mathcal{K}(t)} (P^{\rm{dl}})$, the PS quantizes $\boldsymbol{\theta} (t) - \widehat{\boldsymbol{\theta}} (t_k)$ to a limited rate and broadcasts it to device $k$, $\forall k \in \mathcal{K}(t)$, such that the resultant rate vector is within $\mathcal{C}^{\rm{dl}}_{\mathcal{K}(t)} (P^{\rm{dl}})$.
Accordingly, the PS broadcasts $\boldsymbol{Q} (\boldsymbol{\theta} (t) - \widehat{\boldsymbol{\theta}}_k (t_k), q^{\rm{dl}}_k(t))$ to device $k$, where $q^{\rm{dl}}_k(t)$ is chosen as the largest integer satisfying $R_{q^{\rm{dl}}_k(t)} \le C_{\mathcal{K}(t), k}^{\rm{dl}}$, $\forall k \in \mathcal{K}(t)$.
Therefore, the rate vector $[R_{q^{\rm{dl}}_1(t)}, ..., R_{q^{\rm{dl}}_{K}(t)}]^T$ is within $\mathcal{C}^{\rm{dl}}_{\mathcal{K}(t)} (P^{\rm{dl}})$, and device $k$ can successfully decode $\boldsymbol{Q} (\boldsymbol{\theta} (t) - \widehat{\boldsymbol{\theta}}_k (t_k), q_k(t))$, $\forall k \in \mathcal{K}(t)$.
After receiving $\boldsymbol{Q} (\boldsymbol{\theta} (t) - \widehat{\boldsymbol{\theta}}_k (t_k), q^{\rm{dl}}_k(t))$, device $k$ obtains an estimate of global model $\boldsymbol{\theta} (t)$ through, for $k \in \mathcal{K}(t)$,
\begin{align}
\widehat{\boldsymbol{\theta}}_k (t) = \widehat{\boldsymbol{\theta}}_k (t_k) + \boldsymbol{Q} (\boldsymbol{\theta} (t) - \widehat{\boldsymbol{\theta}}_k (t_k), q^{\rm{dl}}_k(t)),   
\end{align}
which is equivalent to $\widehat{\boldsymbol{\theta}}_k (t) = {\boldsymbol{\theta}} (0) + \sum\nolimits_{i =1}^{t} \mathds{1}(\mathcal{K} (i)) \boldsymbol{Q} \big(\boldsymbol{\theta} (i) - \widehat{\boldsymbol{\theta}}_k (i_k), q^{\rm{dl}}_k(i)\big)$, where we assumed that $\widehat{\boldsymbol{\theta}}_k(0) = \boldsymbol{\theta}(0)$, $\forall k$, and for any time index $i$, we denote the last time device $k$ was selected by $i_k$. 
Accordingly, the PS knows about $\widehat{\boldsymbol{\theta}}_k (t)$ since it has knowledge about $\boldsymbol{Q} (\boldsymbol{\theta} (t) - \widehat{\boldsymbol{\theta}}_k (t_k), q^{\rm{dl}}_k(t))$, $\forall k, t$.

\subsection{Uplink Transmission}
After computing $\widehat{\boldsymbol{\theta}}_k (t)$, device $k \in \mathcal{K}(t)$ performs the $\tau$-step SGD algorithm as given in \eqref{EQTauStepSGD} and obtains $\Delta \boldsymbol{\theta}_k (t)$, which is quantized and transmitted over the MAC through error compensation. 
Specifically, device $k \in \mathcal{K}(t)$ transmits $\boldsymbol{Q}\big( \Delta \boldsymbol{\theta}_k (t) + \boldsymbol{\delta}_k (t), q^{\rm{ul}}_k(t) \big)$, where $q^{\rm{ul}}_k(t)$ is chosen as the largest integer satisfying $R_{q^{\rm{ul}}_k(t)} \le C_{\mathcal{K}(t), k}^{\rm{ul}}$.
The error accumulation vector $\boldsymbol{\delta}_k (t)$ is then updated as
\begin{align}
\boldsymbol{\delta}_k (t+1) = \Delta \boldsymbol{\theta}_k (t) + \boldsymbol{\delta}_k (t) - \boldsymbol{Q}\big( \Delta \boldsymbol{\theta}_k (t) + \boldsymbol{\delta}_k (t), q^{\rm{ul}}_k(t) \big), 
\end{align}
where $\boldsymbol{\delta}_k (0) = \boldsymbol{0}_d$.
After receiving $\boldsymbol{Q}\big( \Delta \boldsymbol{\theta}_k (t) + \boldsymbol{\delta}_k (t), q^{\rm{ul}}_k(t) \big)$, $\forall k \in \mathcal{K}(t)$, the PS updates the global model as
\begin{align}
\boldsymbol{\theta} (t+1) = \sum\nolimits_{m = 1}^M \frac{B_m}{B} \big( \boldsymbol{Q}\big( \Delta \boldsymbol{\theta}_k (t) + \boldsymbol{\delta}_k (t), q^{\rm{ul}}_k(t) \big) + \widehat{\boldsymbol{\theta}}_k (t) \big).
\end{align}

\begin{remark}
By selecting a larger number of devices, each device obtains a coarser estimate of the global model from the PS and sends a less accurate estimate of its local model update to the PS. 
This is due to the resource sharing nature of digital transmission over both downlink and uplink.
On the other hand, having a larger number of devices participating in the training, the global model is updated using a larger portion of the entire dataset.
Therefore, we need to optimize the number of participating devices to obtain the best performance. 
\end{remark}

\begin{remark}
In \cite{mohammadDenizSanjVinceQuanUpdates}, the same version of the global model update with respect to the last model estimate available at the devices is broadcast to all the devices.
This requires that all the devices know about the last global model estimate in order to recover an estimate of the global model.
We have relaxed this requirement by allowing the devices to receive different versions of the global model updates with respect to the last global model estimate available at each device which enables partial device participation.
\end{remark}

\section{Numerical Experiments}\label{SecExperiments}

We evaluate the performance of the proposed algorithm and compare it with the approach in \cite{mohammadDenizSanjVinceQuanUpdates} for image classification of the MNIST dataset \cite{LeCunMNIST} using ADAM optimizer \cite{ADAMDC}.
We train a convolutional neural network with 5 layers including two $5 \times 5$ convolutional layers with ReLU activation and same padding and 32 and 64 channels, respectively, each followed by a $2 \times 2$ max pooling layer, and a softmax output layer.

We assume non-iid data distribution across the devices where each device has access to the samples from only a single class.
We first split the data samples at each class to $M/10$ disjoint groups (assuming that $M$ is divisible by 10), and allocate each group to a unique device.
We measure the performance as the accuracy with respect to the test samples, referred to as the \textit{test accuracy}, versus the iteration count $t$.

We consider $M=100$ devices, $s^{\rm{dl}} = 10^7$ and $s^{\rm{ul}} = 5 \times 10^6$ downlink and uplink sub-channels, respectively, and channel variances $(\sigma^{\rm{dl}})^2 = (\sigma^{\rm{ul}})^2=10$. 
We set the power constraints for downlink and uplink to relatively high values $P^{\rm{dl}} = 10^5$ and $P^{\rm{ul}} = 10^3$, respectively, which is to make sure that in the case of full device participation, i.e., when $K = M$, the PS and the devices can transmit at least one bit of information.

\begin{figure}[t!]
\centering
\centering
\includegraphics[scale=0.575,trim={15.5pt 5pt 44pt 34pt},clip]{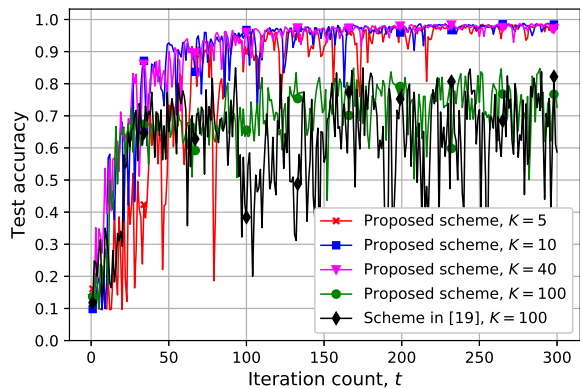}
\caption{Performance comparison of the proposed scheme with different $K$ values and the approach in \cite{mohammadDenizSanjVinceQuanUpdates} for non-iid data.}
\label{Fig}
\end{figure}

In Fig. \ref{Fig}, we investigate the impact of the number of selected devices on the performance for $K \in \{5, 10, 40, 100\}$. 
We observe that selecting all the devices leads to unstable accuracy performance and low convergence rate and accuracy level.
This is due to the fact that the information in both downlink and uplink are communicated less accurately when all the devices participate in the training.
On the other hand, when a small subset of the devices is selected, i.e., $K= 5$, the performance suffers from lack of data usage for updating the global model at each iteration, which in particular deteriorates the performance in the case of non-iid data.
As a result, we observe that selecting a moderate number of devices $K=40$ leads to a relatively good performance in terms of final accuracy level, convergence rate, and stability.
We have further included the performance of the scheme introduced in \cite{mohammadDenizSanjVinceQuanUpdates}, where all the devices are selected at each iteration, and the quantized global model update is broadcast to all the devices with the same rate accommodating the worst device's channel capacity.
We observe that, in the full device participation case, sending the global model update with different rates to different devices based on their channel qualities, as performed by the proposed scheme, can slightly improve the performance.

\section{Conclusions}\label{SecConc}
We have studied FEEL with noisy bandwidth-limited downlink and uplink channels and investigated the advantages of device selection based on downlink channel conditions.  
For the downlink, we have adopted digital transmission of the global model updates from the PS to the devices, where the PS broadcasts different messages to different devices.
Accordingly, we have designed a FEEL framework with partial device participation based on the downlink channel conditions.
The PS broadcasts different global model updates to different participating devices according to the last global model estimate available at the devices.
The participating devices update the global model estimate according to their local data and send the quantized local model updates to the PS, which updates the global model using the local model updates received from the devices. 
Partial device participation enhances the resources allocated to the underlying links and leads to a more accurate information exchange over both the downlink and uplink.
On the other hand, the data usage reduces when fewer devices participate in the training.
Numerical experiments have shown that partial device participation based on the downlink channel improves the performance while the full device participation case suffers significantly from inaccurate information exchange.

\bibliographystyle{IEEEtran}
\bibliography{Report}

\end{document}